\documentclass[conference,10pt]{IEEEtran}
\usepackage{epsfig,epsf,rotating,setspace,latexsym,amsmath,amssymb,amsfonts,bm,theorem,subfigure,epstopdf}
\usepackage{cite,authblk}
\usepackage{bbm}
\usepackage{algorithm}
\usepackage[noend]{algpseudocode}
\usepackage{color}
\usepackage{mathtools}
\usepackage{soul}

\algrenewcommand\algorithmicforall{\textbf{foreach}}
\algrenewcommand\algorithmicindent{.8em}

\newtheorem{theorem}{Theorem}

\newenvironment{Proof}[1]{\medskip\par\noindent{\bf Proof:\,}\,#1}{{\mbox{\,$\blacksquare$}\par}}

\IEEEoverridecommandlockouts
\allowdisplaybreaks

\begin{document}
 
\title{Timely Opportunistic Gossiping in Dense Networks}
 
\author{Purbesh Mitra \qquad Sennur Ulukus\\
        \normalsize Department of Electrical and Computer Engineering\\
        \normalsize University of Maryland, College Park, MD 20742\\
        \normalsize  \emph{pmitra@umd.edu} \qquad \emph{ulukus@umd.edu}}
\maketitle

\begin{abstract}
We consider gossiping in a fully-connected wireless network consisting of $n$ nodes. The network receives Poisson updates from a source, which generates new information. The nodes gossip their available information with the neighboring nodes to maintain network timeliness. In this work, we propose two gossiping schemes, one semi-distributed and the other one fully-distributed. In the semi-distributed scheme, the freshest nodes use pilot signals to interact with the network and gossip with the full available update rate $B$. In the fully-distributed scheme, each node gossips for a fixed amount of time duration with the full update rate $B$. Both schemes achieve $O(1)$ age scaling, and the semi-distributed scheme has the best age performance for any symmetric randomized gossiping policy. We compare the results with the recently proposed ASUMAN scheme \cite{mitra_allerton22}, which also gives $O(1)$ age performance, but the nodes need to be age-aware.
\end{abstract}

\section{Introduction}

Gossiping is an information sharing mechanism where nodes transmit their own data to the neighboring nodes randomly. Gossiping does not require any centralized scheduling and is particularly suitable for communication in dense networks. There are many gossip algorithms in the literature, e.g., \cite{yaron03thesis, shah08monograph, Sanghavi2007GossipFileSplit}, that focus on maximizing the effectiveness of information dispersion. Gossip algorithms have been studied from a \emph{timeliness} point of view in \cite{yates21gossip}. The analysis in \cite{yates21gossip} uses the \textit{version age} metric, which is one of the measures of information freshness in the literature \cite{kaul11AoI, kosta17AoIbook, Sun2019AgeOI, yatesJSACsurvey, cho3BinaryFreshness, zhong18AoSync, maatouk20AOII, melih2020infocom, Abolhassani21version, wang19counting_process, melih_cache_TWT, melih20LimitedCache, kaswan2021ISIT, melih21InfectionTracking, yates21spawc}. 

The analysis in \cite{yates21gossip} shows that for a fully-connected network of $n$ nodes, if each node gossips with a fixed rate $\lambda$, the average version age of any individual node scales as $O(\log n)$ with the network size $n$. Subsequent works show that there can be improvements in the age scaling by introducing particular network mechanisms, such as clustering \cite{buyukates21CommunityStructure, buyukates22ClusterGossip, melih2021globecom}, file slicing and network coding \cite{kaswan22slicingcoding}. Further works consider robustness of timely gossiping against adversarial actions, such as jamming \cite{kaswan22jamming} and timestomping \cite{kaswan22timestomp}, investigate the role of reliable and unreliable sources on the age in gossiping \cite{kaswan23reliable_source}, and study the effects of non-Poisson updating (arbitrary inter-update times) on timeliness in cache-updating tree networks \cite{kaswan23nonpoisson}.

In this paper, we focus on another aspect of existing gossip schemes, which is their uniform gossip rate assignment to all nodes. A main drawback of uniform rate gossiping is that it allocates the same gossip rate to nodes with relatively stale and  relatively fresh information. This negatively impacts the timeliness performance of the network. Our goal in this paper is to efficiently and distributedly allocate the total network gossip capacity dynamically among the users, thereby enabling opportunistic gossiping, where fresher nodes gossip with higher gossip rates.  

The first paper to address the inefficiency of uniform rate gossiping is \cite{mitra_allerton22} which proposed the ASUMAN scheme, which is an opportunistic gossiping scheme that relies on the assumption that the nodes are \emph{age-aware}. In ASUMAN, since the nodes are age-aware, whenever the source updates itself, all the nodes in the network get synchronized and a new gossiping frame starts. When a new frame starts, the nodes stop gossiping and send a small pilot signal after waiting for a back-off period proportional to their current age. In this way, the freshest nodes get to start gossiping first as their back-off period is smallest and the relatively staler nodes do not gossip after receiving the pilot signal from the freshest nodes. If in any frame, the number of fresh nodes is more than one, then that number is estimated from the received pilot signals and the total update rate $B=n\lambda$ is equally divided between them. The analysis in \cite{mitra_allerton22} shows that the version age of an individual node scales as $O(1)$ with the network size $n$. 

Although ASUMAN achieves better age performance, the system model poses some challenges in real-life implementations. One such challenge is that when multiple nodes have the same minimum age, all of them transmit the pilot signal simultaneously. Thus, multiple short signals overlap over the air, which leads to incorrect estimation of the minimum-age nodes, causing interference within the gossiping nodes. Another downside of ASUMAN is that the nodes have to be age-aware. This can be achieved if the source sends a signal to the nodes when it updates itself, adding additional complexity to the simple gossiping model.

\begin{table}[h]
    \begin{center}
    \begin{tabular}{|l|l|}
        \hline
        \textbf{gossiping scheme}                     & \textbf{age scaling} \\ \hline 
        ASUMAN proposed in \cite{mitra_allerton22}    & $2\frac{\lambda_e}{\lambda}+1$ \\ \hline
        semi-distributed proposed here \qquad \qquad  & $2\frac{\lambda_e}{\lambda}$ \\ \hline
        fully-distributed proposed here               & $(1+e)\frac{\lambda_e}{\lambda}$ \\ \hline 
    \end{tabular}
    \end{center}
    \caption{Age scaling comparison for different gossiping schemes.}
    \label{age_scale_comp}
    \vspace*{-0.4cm}
\end{table}

In this paper, we propose two new gossiping schemes, one semi-distributed and the other fully-distributed, that both yield $O(1)$ performance. These schemes are able to circumvent the previously mentioned downsides. In the semi-distributed scheme, each time a node gets updated by the source, it transmits a pilot signal to the neighboring nodes and starts gossiping with the maximum capacity until it receives a signal from some other node. In the fully-distributed scheme, each time a node gets updated by the source, it gossips for a fixed duration with the maximum capacity and stops. The age scaling comparison of these schemes is shown in Table~\ref{age_scale_comp}. Further, we prove that the semi-distributed gossiping scheme yields the best age performance among all possible symmetric gossiping schemes with an upper bound on the instantaneous maximum gossip rate. For our analysis, we use stochastic hybrid system (SHS) formulation \cite{hespanha_SHS}, similar to \cite{yates21gossip, mitra_allerton22}, to calculate of mean steady-state version age of the nodes.

\section{System Model}

\begin{figure}[t]
\centerline{\includegraphics{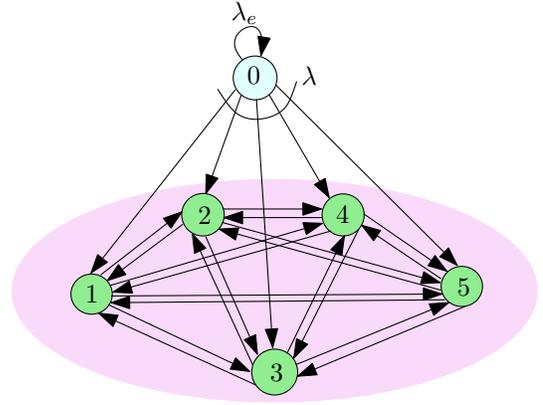}}
\caption{Source $0$ updates itself with rate $\lambda_{e}$ and sends updates to the nodes $\mathcal{N}=\{1,2,3,4,5\}$ uniformly with total rate $\lambda$, i.e., with rate $\lambda/5$ to each of the nodes. The nodes gossip with each other with total update rate $B$.}
\label{SystemModel}
\end{figure}

We consider a gossip network consisting of a source labeled node $0$, and a set of nodes labeled $\mathcal{N}=\{1,2,\ldots,n\}$, as shown in Fig.~\ref{SystemModel}. The source updates its information with Poisson arrivals of rate $\lambda_e$, and it sends Poisson updates to the network with a total rate of $\lambda$. For simplicity, we consider a symmetric network, i.e., each of the nodes receives updates from the source with a rate $\frac{\lambda}{n}$. In the timely gossiping papers in the literature \cite{yates21gossip, yates21spawc,buyukates21CommunityStructure, buyukates22ClusterGossip, melih2021globecom, kaswan22slicingcoding, kaswan22jamming, kaswan22timestomp}, it is assumed that each node of the network gossips with a rate of $\lambda$; thus, on a fully connected network where each node is connected to $(n-1)$ other nodes, each node $i$ gossips with a node $j$ with a rate of $\frac{\lambda}{n-1}$. Therefore, the total update capacity of the network is $B=n \lambda$. As in \cite{mitra_allerton22}, in this paper, we consider allocating this total update rate $B$ to users dynamically. Once a gossip rate is assigned to a node, it gossips with its $(n-1)$ neighbors with equal rates in the fully connected network.

Thus, the network has an upper bound of $B$ on the instantaneous gossiping rate. If at any time, multiple nodes transmit and the total instantaneous gossip rate exceeds $B$, there will be interference, and the gossiped data is lost. Hence, for effective gossiping, at any time instant, the total instantaneous gossip rate has to be less than or equal to $B$. The goal of our work is to improve the timeliness of such a network. To measure the timeliness of the $i$th node, we use version age, denoted as $\Delta_i(t)$. This measure counts how many versions the data at the $i$th node is lagging, compared to the data available at the source at time $t$. Mathematically, we write
\begin{align}
    \Delta_i(t)=N_s(t)-N_i(t),
\end{align}
where $N_s(t)$ and $N_i(t)$ are the versions of the data available at the source and at the $i$th node, respectively, at time $t$. We denote all the ages of nodes at time $t$ as the age vector $\mathbf{\Delta}(t)=[\Delta_1(t),\Delta_2(t),\ldots,\Delta_n(t)]$. When the source updates itself, all the ages of the nodes increase by $1$. If the source sends an update to a node, its age becomes $0$. When node $i$ sends a gossip update to node $j$, it stores the data with the freshest version, i.e., the age of the $j$th node becomes $\hat{\Delta}_j(t)=\Delta_{\{i,j\}}(t)=\min\{\Delta_i(t),\Delta_j(t)\}$.

\section{Semi-Distributed Gossiping}

In this section, we introduce the semi-distributed gossiping scheme. The motivation for this is to allow the freshest node of the network to gossip with maximum capacity. Suppose we denote the $k$th source-to-$i$th node update as $t^{(i)}_{k}$. In this scheme, at time $t^{(i)}_k$, node $i$ transmits a small pilot signal to all the other nodes in the network and starts gossiping with rate $B$ to the other nodes with equal rate. While gossiping, if node $i$ receives a pilot signal from any other node, it will stop gossiping. We define the gossiping node at any given time $t$ as $\mathcal{M}(t)$. Since, the probability of two simultaneous Poisson arrivals is $0$, i.e., $\mathbb{P}(|\mathcal{M}(t)|\geq 2)=0$, here we do not face the problem of overlapping pilot signals like ASUMAN\cite{mitra_allerton22}.

We investigate the mean steady-state age of an individual node, denoted as
\begin{align}
    a_i=\lim_{t\to\infty}a_i(t)=\lim_{t\to\infty}\mathbb{E}[\Delta_i(t)],
\end{align}
in particular, how network size $n$ affects $a_i$, in Theorem~\ref{semi_dist_thm}.

\begin{theorem}\label{semi_dist_thm}
If $B=n\lambda$, the average version age of a node $a_i$ in a semi-distributed gossip network scales as $O(1)$.
\end{theorem}

\begin{Proof}
We use SHS formulation of \cite{hespanha_SHS}. Note that, for any time $t$, the gossiping node is the minimum age node in the network. Let us denote this minimum age as $\Delta_{\min}(t)=\min\{\Delta_1(t),\Delta_2(t),\ldots,\Delta_n(t)\}$. From \cite{yates21gossip}, we know that $\lim_{t\to\infty}\mathbb{E}[\Delta_{\min}(t)]=\frac{\lambda_e}{\lambda}$. Since for any given $t$, only the node with the minimum age is gossiping, we can express the state transition of the system as an SHS with only one type of transition, i.e., $\mathcal{Q}=0$. We choose the test function $\psi_i:\mathbbm{R}^n\times[0,\infty)\to\mathbbm{R}$, where $i\in\mathcal{N}$, as 
\begin{align}
    \psi_i(\mathbf{\Delta}(t),t)=\Delta_i(t). 
\end{align}
Now, following \cite[Thm.~1]{hespanha_SHS}, we evaluate the extended generator function as
\begin{align}\label{extended_gen}
    \mathbb{E}[(L\psi_i)(\mathbf{\Delta}(t),t)]=&\sum_{(j,\ell)\in \mathcal{L}}\lambda_{j,\ell}(\mathbf{\Delta}(t),t)\mathbb{E}\big[\psi_i(\phi_{j,\ell}(\mathbf{\Delta}(t),t))\notag\\
    &\qquad-\psi_i(\mathbf{\Delta}(t),t)\big],
\end{align}
where $\mathcal{L}$ denotes all possible state transitions. We define the reset maps $\phi_{j,\ell}(\mathbf{\Delta}(t),t)=\hat{\mathbf{\Delta}}(t)=[\hat{\Delta}_1(t),\hat{\Delta}_2(t),\ldots,\hat{\Delta}_n(t)]$ as follows
\begin{align}
    \hat{\Delta}_i(t)=
    \left\{
	\begin{array}{ll}
		\Delta_i(t)+1,  &\mbox{if } j = 0, \ell=0 \\
		0, &\mbox{if } j=0, \ell=i\\
		\min(\Delta_j(t),\Delta_\ell(t)),  & \mbox{if } j\in\mathcal{N},\ell=i \\
		\Delta_i(t), &\mbox{otherwise}.
	\end{array}
\right.
\end{align}
The update rates $\lambda_{j,\ell}$ are given as
\begin{align}\label{update_rates}
    \!\!\lambda_{j,\ell}(\mathbf{\Delta}(t),t)=
    \left\{
	\begin{array}{ll}
		\lambda_{e}, &\mbox{if } j = 0,\ell=0 \\
		\frac{\lambda}{n}, &\mbox{if } j=0,\ell=i\\
		\frac{B}{n-1}\mathbbm{1}\{j=\mathcal{M}(t)\}, &\mbox{otherwise},
	\end{array}
\right.
\end{align}
where $\mathbbm{1}\{\cdot\}$ denotes the indicator function. Now, we can rewrite (\ref{extended_gen}) as
\begin{align}\label{extended_gen_expn}
   \mathbb{E}&[(L\psi_{i})(\mathbf{\Delta}(t),t)]\notag\\
    &=\mathbb{E}\bigg[\lambda_{e}(\Delta_i(t)+1-\Delta_i(t))+\frac{\lambda}{n}(0-\Delta_i(t))\notag\\
    &\quad \ \ +\sum_{j\in \mathcal{N}}\frac{B}{n-1}\mathbbm{1}\{j=\mathcal{M}(t)\}\left(\Delta_{\{j,i\}}(t)-\Delta_i(t)\right)\bigg].
\end{align}
Since the gossiping node is always the minimum age node, we can write
\begin{align}
  \mathbb{E}&[(L\psi_{i})(\mathbf{\Delta}(t),t)]\notag\\
    &=\lambda_{e}-\frac{\lambda}{n}a_i(t)+\mathbb{E}\bigg[\sum_{j= \mathcal{M}(t)}\frac{B}{n-1}(\Delta_{\min}(t)-\Delta_i(t))\bigg]\notag\\
    &=\lambda_{e}-\frac{\lambda}{n}a_i(t)+\frac{B}{n-1}(a_{\min}(t)-a_i(t)).\label{GossipEquation}
\end{align}
Now, since the version age is a piece-wise constant function of time, we obtain
\begin{align}
    \frac{d\mathbb{E}[\psi_i(\mathbf{\Delta}(t),t)]}{dt}=\frac{d\mathbb{E}[\Delta_i(t)]}{dt}=0,
\end{align}
for any continuity point $t$. Hence, the expected value in \eqref{GossipEquation} is $0$, by Dynkin’s formula, as given in \cite{hespanha_SHS}. Thus, \eqref{GossipEquation} becomes
\begin{align}
    0=\lambda_{e}-\frac{\lambda}{n}a_i(t)+\frac{B}{n-1}(a_{\min}(t)-a_i(t)).
\end{align}
Hence, the mean age of an individual node is expressed as
\begin{align}\label{a_i_t}
    a_i(t)=\frac{\lambda_e+\frac{B}{n-1}a_{\min}(t)}{\frac{\lambda}{n}+\frac{B}{n-1}}.
\end{align}
To evaluate the steady-state mean age, we take $t\to\infty$ in (\ref{a_i_t}) which gives
\begin{align}\label{mean_age_eqn}
    a_i=\frac{\lambda_e+\frac{B}{n-1}\frac{\lambda_e}{\lambda}}{\frac{\lambda}{n}+\frac{B}{n-1}}.
\end{align}
Finally, to calculate the scaling of the average age, we use $B=n\lambda$, which yields
\begin{align} \label{mean_age_eqn_lim}
    \lim_{n\to\infty}a_i=\lim_{n\to\infty}\frac{\lambda_e}{\lambda}\left(\frac{1+\frac{n}{n-1}}{\frac{1}{n}+\frac{n}{n-1}}\right)=\frac{2\lambda_{e}}{\lambda},
\end{align}
concluding the proof.
\end{Proof}

Next, we show that this semi-distributed scheme gives the best version age performance for any possible gossiping scheme with a constraint on the instantaneous gossiping scheme, in Theorem~\ref{lower_bound_thm}. 

\begin{theorem}\label{lower_bound_thm}
For any symmetric network with maximum instantaneous gossip rate of $B$, the semi-distributed gossiping scheme yields the minimum average age for the nodes.
\end{theorem}

\begin{Proof}
Suppose we use any arbitrary gossiping policy. Since the total gossip rate is upper bounded by $B$, we have
\begin{align}\label{update_rate_bound1}
    \sum_{j,i\in\mathcal{N},j \neq i}\lambda_{j,i}(\mathbf{\Delta}(t),t)\leq B, \quad \forall t.
\end{align}
From the symmetry of the network, we can write
\begin{align}\label{update_rate_bound}
\mathbb{E}\left[\sum_{j\in\mathcal{N},j \neq i}\lambda_{j,i}(\mathbf{\Delta}(t),t)\right]\leq\frac{B}{n-1}.
\end{align}
Note that the sum in (\ref{update_rate_bound1}) is over all $i,j$ whereas the sum in  (\ref{update_rate_bound}) is over $j$ only. Now, equating the extended generator function to $0$, yields
\begin{align}\label{lower_bound_equation}
    \frac{\lambda}{n}a_i(t)&+\mathbb{E}\left[\sum_{j\in\mathcal{N},j \neq i}\lambda_{j,i}(\mathbf{\Delta}(t),t)\Delta_i(t)\right]\notag\\   &=\lambda_e+\mathbb{E}\left[\sum_{j\in\mathcal{N},j \neq i}\lambda_{j,i}(\mathbf{\Delta}(t),t)\Delta_{\{j,i\}}(t)\right].
\end{align}
Using the inequality in (\ref{update_rate_bound}) and by definition the fact that $\Delta_{\{j,i\}}(t)\geq\Delta_{\min}(t)$, we can rewrite (\ref{lower_bound_equation}) as 
\begin{align}\label{lower_bound_equation_2}
    \frac{\lambda}{n}a_i(t)+\frac{B}{n-1}a_i(t)\geq\lambda_e+\frac{B}{n-1}a_{\min}(t).
\end{align}
Taking $t\to\infty$ in (\ref{lower_bound_equation_2}) and using the expression of $a_{\min}(t)$, we obtain
\begin{align}
    a_i\geq\frac{\lambda_e+\frac{B}{n-1}\frac{\lambda_e}{\lambda}}{\frac{\lambda}{n}+\frac{B}{n-1}},
\end{align}
where the right-hand side of the inequality is the average age of a node with the proposed semi-distributed policy. This concludes the proof.
\end{Proof}

\begin{figure*}[t]
\centerline{\includegraphics[width=\textwidth]{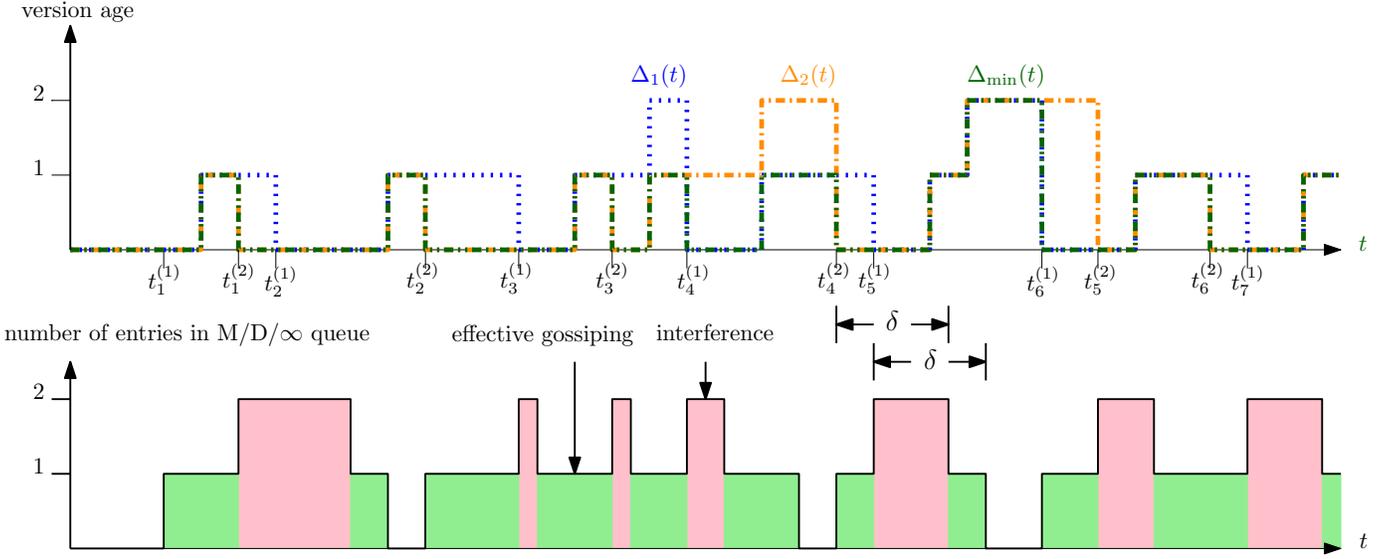}}
\caption{Distributed gossiping in a 2 node network. At each $t^{(i)}_k$, $\Delta_i(t)$ becomes zero and node $i$ starts gossiping for a $\delta$ duration. The corresponding M/D/$\infty$ queue indicates the number of nodes gossiping simultaneously. Effective gossiping only happens when only one node is gossiping. Presence of multiple gossiping nodes creates interference, resulting in no net gossip.}
\label{age_evol}
\end{figure*}

\section{Fully-Distributed Gossiping}

In this section, we introduce a gossiping policy which is fully-distributed. In ASUMAN \cite{mitra_allerton22}, the nodes need to be age-aware and in the semi-distributed scheme, the nodes need to implement a pilot-signal based communication in the network. We improve upon them and formulate a gossiping policy that does not require age-awareness or pilot-signal transmissions. In this scheme, whenever node $i$ receives an update from the source at time $t^{(i)}_k$, it starts gossiping to all the other nodes with rate $B$ for a fixed time duration $\delta$, and then it stops, as shown in Fig.~\ref{age_evol}. We investigate the age performance of this scheme in Theorem~\ref{dist_thm}.

\begin{theorem}\label{dist_thm}
If $B=n\lambda$, the average version age of a node in a fully-distributed gossip network scales as $O(1)$.
\end{theorem}

\begin{Proof}
From Fig.~\ref{age_evol}, we observe that at any given time, if there is any effective gossiping, only the minimum age node is responsible for it. This is because, effective gossiping is possible only if a single node is gossiping and in that case, the node has to be a minimum age node. Whereas, when multiple nodes are gossiping with rate $B$, there will be no effective gossiping due to interference. Additionally, each update from the source is a Poisson arrival with rate $\lambda$, and gossiping starts immediately for a time duration of $\delta$. Hence, this system is equivalent to an M/D/$\infty$ queue. Now, from \cite{Bolch1998QueueingNA,Newell_queue}, we know that the stationary distribution for any general M/G/$\infty$ queue follows the Poisson distribution,
\begin{align}
    \pi_k=\frac{(\lambda/\mu)^k e^{-\lambda/\mu}}{k!}, \quad k=0,1,2,\ldots,
\end{align}
where $\pi_k$ is the stationary probability of having $k$ entries in the queue. For this M/D/$\infty$ queue, $\mu=\frac{1}{\delta}$. Since effective gossip happens only when there is one entry in the queue, the effective gossip rate becomes 
\begin{align}
    \tilde{B}=\pi_1 B=\lambda\delta e^{-\lambda\delta}B.
\end{align}
The rest of the analysis is the same as in Theorem~\ref{semi_dist_thm}. Therefore, we can directly substitute $\tilde{B}$ instead of $B$ in (\ref{mean_age_eqn}) to obtain the mean age of the $i$th node as
\begin{align}\label{mean_age_distb}
    a_i=\frac{\lambda_e+\frac{\tilde{B}}{n-1}\frac{\lambda_e}{\lambda}}{\frac{\lambda}{n}+\frac{\tilde{B}}{n-1}}.
\end{align}
Using $B=n\lambda$ and taking $n\to\infty$ in (\ref{mean_age_distb}), we get the age scaling as
\begin{align}
    \lim_{n\to\infty}a_i&=\lim_{n\to\infty}\frac{\lambda_e+\frac{\lambda\delta e^{-\lambda\delta}n\lambda}{n-1}\frac{\lambda_e}{\lambda}}{\frac{\lambda}{n}+\frac{\lambda\delta e^{-\lambda\delta}n\lambda}{n-1}}\\
    &=\frac{\lambda_e}{\lambda}\left(1+\frac{1}{\lambda\delta e^{-\lambda\delta}}\right), \label{mean_age_scale_distb}
\end{align}
which concludes the proof.
\end{Proof}

Finally, we note that the age expression in (\ref{mean_age_scale_distb}) for the fully-distributed gossiping scheme depends on the chosen gossiping duration $\delta$. Thus, we can improve the age expression in (\ref{mean_age_scale_distb}) by choosing an optimal $\delta$ that minimizes the mean age. Since $\lambda\delta e^{-\lambda\delta}\leq \frac{1}{e}$, the maxima being at $\delta^*=\frac{1}{\lambda}$, the lower bound of mean age of distributed gossiping is $\frac{\lambda_e}{\lambda}\left(1+\frac{1}{e^{-1}}\right)=(1+e)\frac{\lambda_e}{\lambda}$. This result matches our intuition, because if $\delta$ is too small, it will not allow sufficient time to gossip. On the other hand, if $\delta$ is too large, there will not be effective gossiping due to interference from simultaneous gossiping nodes. The minimum age is achieved when the effective gossiping rate $\tilde{B}$ is maximized, which is $\tilde{B}|_{\delta^*}=\frac{B}{e}$.

\begin{figure}[p]
\subfigure[$\frac{\lambda_e}{\lambda}=0.4$]{
\centering
\includegraphics[scale=0.47]{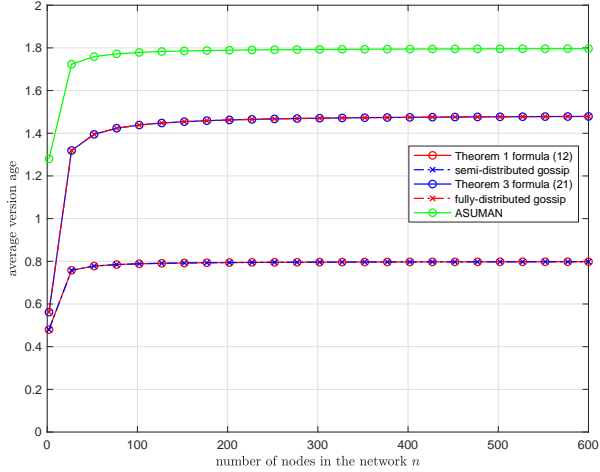}
\label{dist_vs_semidist_r.4}
}
\subfigure[$\frac{\lambda_e}{\lambda}=1$]{
\centering
\includegraphics[scale=0.47]{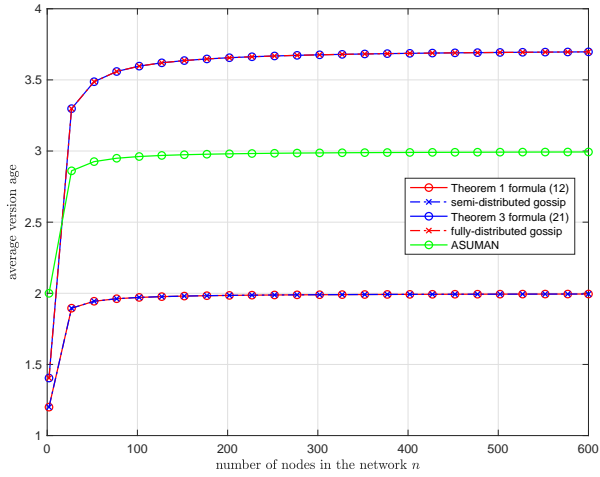}
\label{dist_vs_semidist_r1}
}
\subfigure[$\frac{\lambda_e}{\lambda}=2$]{
\centering
\includegraphics[scale=0.47]{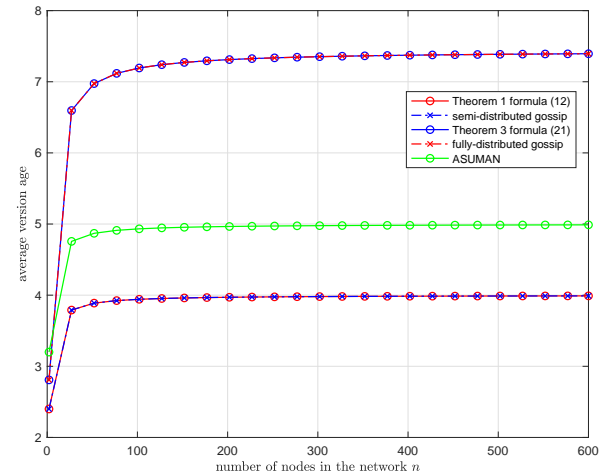}
\label{dist_vs_semidist_r2}
}
\caption{Average version age of a single node versus the total number of nodes in the network $n$ for semi-distributed, fully-distributed and ASUMAN schemes.}
\label{dist_vs_semidist}
\end{figure}

\section{Numerical Results}

In this section, we present simulation results for the two proposed gossiping schemes, and compare them with the theoretically derived age expressions. We also show the results for ASUMAN \cite{mitra_allerton22} as a benchmark.

In Fig.~\ref{dist_vs_semidist}, we present the numerical results for $\frac{\lambda_e}{\lambda}=0.4$, $\frac{\lambda_e}{\lambda}=1$ and $\frac{\lambda_e}{\lambda}=2$ in
Fig.~\ref{dist_vs_semidist_r.4}, Fig.~\ref{dist_vs_semidist_r1} and Fig.~\ref{dist_vs_semidist_r2}, respectively, with $\lambda=1$ in all cases. From the figures, it is evident that all the gossiping schemes result in $O(1)$ performance and the semi-distributed gossiping scheme yields the best performance among all. 

In Fig.~\ref{dist_vs_semidist_r.4}, where $\frac{\lambda_e}{\lambda}=0.4<\frac{1}{e-1}$, ASUMAN gives the worst age performance among the three schemes. However, in Fig.~\ref{dist_vs_semidist_r1} and Fig.~\ref{dist_vs_semidist_r2}, i.e., for $\frac{\lambda_e}{\lambda}>\frac{1}{e-1}$, ASUMAN performs worse than the semi-distributed scheme, but is better than the fully-distributed scheme. This matches our intuition because, in ASUMAN, we use the information about source self-updates to allocate gossip rate more efficiently, while in the fully-distributed scheme, multiple nodes gossiping together causes interference to lose some portion of the total gossip rate. This effect of interference becomes more prominent when the source to network update rate $\lambda$ is high as compared to source self-update rate $\lambda_e$. We have chosen $\delta=\frac{1}{\lambda}=1$ for the simulation to get the minimum average age for fully-distributed gossiping. 

For ASUMAN, the asymptotic age scales as $\lim_{n\to\infty}\frac{\lambda_e}{\lambda}\left(\frac{1+\frac{n}{n-1}(1+\frac{\lambda}{\lambda_e})}{\frac{1}{n}+\frac{n}{n-1}}\right)=2\frac{\lambda_e}{\lambda}+1$, while the other two schemes obtain $2\frac{\lambda_e}{\lambda}$ and $(1+e)\frac{\lambda_e}{\lambda}$, as shown in (\ref{mean_age_eqn_lim}) and (\ref{mean_age_scale_distb}) (with optimized $\delta$), respectively, and as listed in Table~\ref{age_scale_comp}. The numerical simulation results exactly match the derived formulas. With an increase in the ratio $\frac{\lambda_e}{\lambda}$, the average age increases due to source being updated more frequently compared to the network for all schemes, as we observe going from Fig.~\ref{dist_vs_semidist_r.4} to Fig.~\ref{dist_vs_semidist_r1} to Fig.~\ref{dist_vs_semidist_r2}.

\section{Conclusion}
We proposed a semi-distributed and a fully-distributed gossiping scheme for a fully-connected network. The semi-distributed scheme allows the freshest node to communicate in the network through pilot signals and to gossip with full capacity. This scheme archives the lowest possible average age for any symmetric network, with a constraint on the instantaneous gossip rate. On the other hand, in the fully-distributed scheme, the freshest node gossips for a fixed time duration with full capacity. The effective gossip happens only a fraction of the total time, when there is no interference from multiple nodes gossiping. Both of the proposed schemes yield $O(1)$ age performance. Compared to our previous work ASUMAN, which also gives $O(1)$ age scaling, this work is an improvement because here we do not require the nodes to be age-aware or to transmit pilot signals for channel reservation.

\bibliographystyle{unsrt}
\bibliography{reference}

\end{document}